\documentclass[runningheads]{llncs}
\usepackage[T1]{fontenc}
\usepackage{graphicx,verbatim}

\usepackage{booktabs}   % For professional-quality rules
\usepackage{diagbox}    % For diagonal header
\usepackage[table]{xcolor} % For light coloring
\usepackage{array}       % For better column control
\usepackage[table]{xcolor}
\usepackage{colortbl}
\usepackage{makecell}
\definecolor{customcolor1}{RGB}{192,227,223}
\usepackage{multirow}    % for \multirow
\usepackage{array}       % for >{\centering\arraybackslash} if needed
\usepackage{algorithm}
\usepackage{algpseudocode}
\usepackage{amsmath}
\usepackage{subcaption}
\usepackage{amssymb}
\usepackage[colorlinks=true, allcolors=blue, breaklinks=true]{hyperref}
\usepackage{caption}
\usepackage{subcaption}
\begin{document}
% %

\title{Edge2Prompt: Modality-Agnostic Model for Out-of-Distribution Liver Segmentation}

\author{Nathan Hollet \and Oumeymah Cherkaoui \and Philippe C. Cattin \and Sidaty El hadramy}
\authorrunning{Hollet et al.}
\institute{Department of Biomedical Engineering, University of Basel, Allschwil, \\ Switzerland. \\ 
\email{n.hollet@stud.unibas.ch}
\email{\{oumeymah.cherkaoui,philippe.cattin,sidaty.elhadramy\}@unibas.ch}}
\maketitle              % typeset the header of the contribution
\begin{abstract}
Liver segmentation is essential for preoperative planning in interventions like tumor resection or transplantation, but implementation in clinical workflows faces challenges due to modality-specific tools and data scarcity. We propose \textbf{Edge2Prompt}, a novel pipeline for modality-agnostic liver segmentation that generalizes to out-of-distribution (OOD) data. Our method integrates classical edge detection with foundation models. Mo\-dality-agnostic edge maps are first extracted from input images, then processed by a U-Net to generate logit-based prompts. These prompts condition the Segment Anything Model 2 (SAM-2) to generate 2D liver segmentations, which can then be reconstructed into 3D volumes. Evaluated on the multi-modal CHAOS dataset, \textbf{Edge2Prompt} achieves competitive results compared to classical segmentation methods when trained and tested in-distribution (ID), and outperforms them in data-scarce scenarios due to the SAM-2 module. Furthermore, it achieves a mean Dice Score of \textbf{86.4\%} on OOD tasks, outperforming U-Net baselines by \textbf{27.4\%} and other self-prompting methods by \textbf{9.1\%}, demonstrating its effectiveness. This work bridges classical and foundation models for clinically adaptable, data-efficient segmentation. The source code will be made publicly available.

\keywords{Liver Segmentation \and Modality-Agnostic Model \and Out-of-Distribution Segmentation \and SAM-2}
% Authors must provide keywords and are not allowed to remove this Keyword section.

\end{abstract}
\section{Introduction}
Effective preoperative planning is essential for liver interventions, including surgeries like tumor removal or liver transplantation, and radiotherapy treatments \cite{horkaew2023recent}. Currently, clinicians rely on 3D abdominal scans \cite{sakas2002trends} (such as MRI or CT) taken before the procedure. To plan effective treatment, it is crucial to segment the liver from the surrounding organs to facilitate accurate quantitative analysis of its shape, boundaries, and volume. While MRI's non-invasive nature and superior soft tissue contrast make it ideal for many patients \cite{noureddin2015mri}, its high cost per scan and scanner installation limit its widespread availability, leading hospitals to often rely on CT-based imaging modalities \cite{geethanath2019accessible}. This creates a practical challenge: hospitals with only CT cannot make use of MRI-specific segmentation tools, and even centers possessing both scanners can be limited by non-robust tools. Therefore, a modality-agnostic segmentation pipeline capable of out-of-distribution (OOD) generalization and accurate 3D volume reconstruction, while requiring minimal training data, is clinically essential.

The pursuit of efficient image segmentation has long driven the development of fundamental techniques \cite{kaur2014various}, including edge-based approaches \cite{Sun2022-xj}, thresholding methods \cite{Amiriebrahimabadi2024-ow}, and clustering techniques \cite{Gangwar2015-ry}. These methods offer a clear advantage over manual segmentation, which is both time-consuming and requires significant expert knowledge. However, their slow inference time and reliance on predefined rules limit their adaptability to complex medical images \cite{xu2024advances}, leading to gradual replacement by deep learning (DL) methods.

While DL based methods demonstrate potential for efficient and adaptable medical image segmentation \cite{SUBASI2024377,El_hadramy2023-jb}, their clinical adoption faces two persistent barriers: modality specificity and dependency on scarce expert annotations. The widely adopted U-Net architecture \cite{DBLP:journals/corr/RonnebergerFB15} exemplifies these limitations. Without extensive multi-modal training data, U-Net remains constrained to specific modalities and susceptible to intra-modality variations (e.g., contrast, noise, resolution differences across MRI protocols). This sensitivity induces domain shifts \cite{yan2019domain}, degrading performance on unseen data and limiting clinical utility.

Foundation models address medical imaging's data scarcity challenge \cite{lee2024foundation} by enabling robust multi-modal segmentation through large-scale pretraining \cite{azad2023foundationalmodelsmedicalimaging}. Their zero-shot capabilities facilitate training-free applications. The current state-of-the-art in this area is SAM-2, whose impressive natural image segmentation, achieved via pretraining on the SA-V dataset \cite{ravi2024sam2}, has spurred significant medical adaptation research. However, direct use in a medical context faces limitations: processing of 2D images only (vs. 3D volumes) and degraded performance on challenging features such as low contrast, irregular shapes, weak boundaries, and small objects \cite{zhang2024segment,mazurowski2023segment}. Consequently, adaptation efforts, typically fine-tuning components \cite{ma2024segment,Gu_2025} or adding domain modules \cite{wu2025medical}, aim to achieve competitive performance against task-specific models. However, fundamental limitations persist. MedSAM \cite{ma2024segment} demonstrates significant modality bias toward MRI, CT, and endoscopy, requiring full fine-tuning of the original SAM model. \textit{Wu et al.} \cite{wu2025medical} circumvent this via a medical-knowledge adapter, improving in-domain quality to clinically acceptable levels but failing to achieve domain independence. Additionally, SAM's reliance on manual prompting represents another key drawback, though \textit{Wu et al.} \cite{wu2023self} have introduced self-prompting via logistic regression to automate coarse mask generation and prompt sampling. Despite these limitations, such extensions underscore SAM's potential for integration into clinically effective medical segmentation pipelines.

In this paper, we propose \textbf{Edge2Prompt}, a novel pipeline for automatic, data-efficient liver segmentation from OOD imaging modalities, coupled with coherent 3D volume reconstruction. Our method uniquely integrates classical and modern techniques: edge maps are forwarded to a U-Net to generate high-quality modality-agnostic logits. These logits then prompt SAM-2 to refine segmentation despite minimal training data. Finally, we reconstruct the segmented slices into a coherent 3D volume.

\section{Methodology}
\begin{figure}[h]
    \centering
    \includegraphics[width=1.0\linewidth]{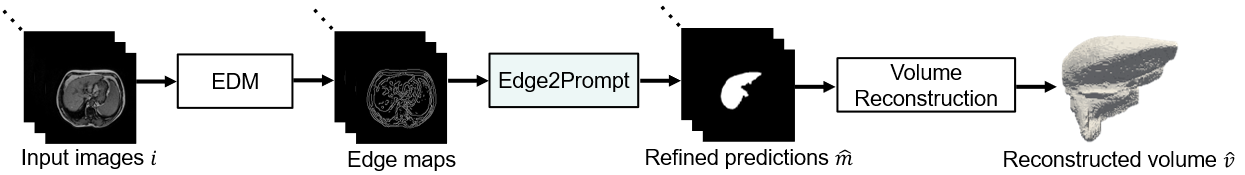}
    \caption{General workflow: The Edge Detection Module (EDM) extracts modality-agnostic edge maps from input images, which the \textbf{Edge2Prompt} block processes to generate refined segmentation mask predictions, enabling 3D organ model reconstruction from segmented slices.}
    \label{fig:general}
\end{figure}

Let the total training dataset consist of $n$ samples indexed by $T = \{1, 2, \dots, n\}  \\ \subseteq \mathbb{N}$. For each $t \in T$, $i_t$ and $m_t$ respectively represent the image and its corresponding ground truth mask. The training dataset is defined as: $D_T^{TRAIN} = \left\{ (i_t, m_t) \mid t \in T \right\}$ and the sub-dataset as: $D_J^{TRAIN} = \left\{ (i_j, m_j) \mid j \in J \right\}$ for any subset $J \subseteq T$. This work aims to develop a plug-and-play segmentation pipeline that produces a segmented liver volume from a series of 2D abdominal scans of a patient. To achieve robust segmentation under limited training data $D_J$ while preserving cross-modality generalization, we introduce a hybrid architecture trained end-to-end and called \textbf{Edge2Prompt}, followed by volumetric reconstruction of the segmented organ. This approach enables accurate OOD image segmentation and subsequent 3D organ reconstruction while maintaining modality-agnostic properties. The pipeline is illustrated in \textbf{Fig. \ref{fig:general}}.
 
\subsection{From 2D CT/MR Images to Edge Maps}
The Edge Detection Module (EDM) first preprocesses the input grayscale image through histogram equalization and bilinear filtering, then extracts edge maps from the resulting 2D image $i \in \mathbb{R}^{240\times 240}$ using a Canny filter ($C$) \cite{Sun2022-xj}. We threshold the Canny response $C(i) \in \mathbb{R}^{240 \times 240}$ to obtain a binary magnitude map. This binarization reduces the impact of modality-specific intensity variations while preserving essential topological structures.

\subsection{From Edges to Segmentation Masks}
\begin{figure}[h]
    \centering
    \includegraphics[width=1.0\linewidth]{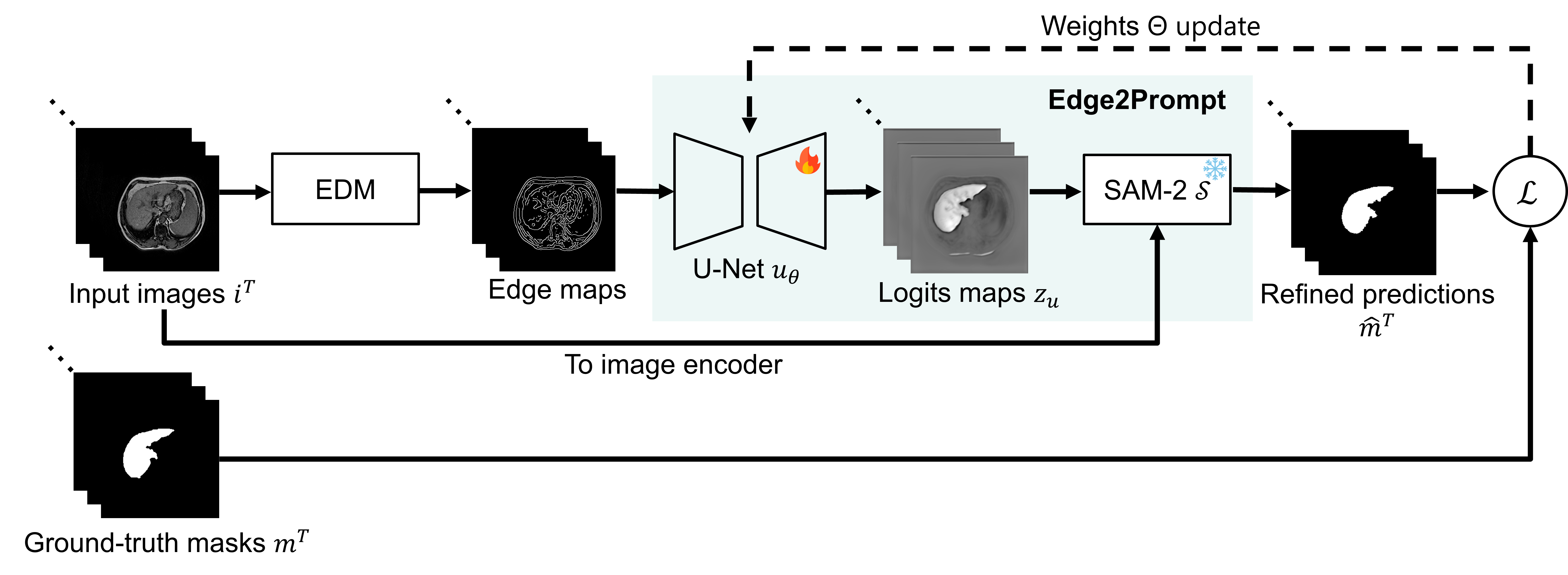}
    \caption{\textbf{Edge2Prompt} training workflow: Modality-agnostic edge maps generated by the EDM serve as input to a modified U-Net architecture. This U-Net processes the edge information to generate a logit-based prompt that conditions a frozen SAM-2 module. The prompted SAM-2 module then generates refined segmentation mask predictions. Using a composite loss function, these predictions are then evaluated against ground-truth masks. Symbol key: flame icon (learnable U-Net parameters), snowflake icon (frozen SAM-2 weights).}
    \label{fig:training}
\end{figure}

The edge representations from the previous subsection serve as input to our \textbf{Edge2Prompt} pipeline. Central to this framework is a U-Net that transforms edge maps into logit-based prompts for conditioning SAM-2. We now detail \textbf{Edge2Prompt} architecture and its end-to-end training logic (\textbf{Fig. \ref{fig:training}}). \\

\noindent \textbf{Prompt Generator}
The chosen architecture for this task is a slightly modified U-Net \cite{DBLP:journals/corr/RonnebergerFB15} implementation using a bilinear upsampling block followed by a convolution for the decoder steps. It takes as input the previously presented edge maps tensor and generates a 2D logits map $z_u = u_\theta(EDM(i)) \in \mathbb{R}^{240 \times 240}$ used to prompt the SAM-2 module. \\

\noindent \textbf{SAM-2 Wrapper}
To backpropagate through the pipeline and update the U-Net weights $\theta$, we designed a differentiable SAM-2 wrapper ($\mathcal{S}$). This implementation forwards the U-Net's preprocessed logits map to SAM-2's prompt encoder, while separately processing the image through the image encoder. Throughout training, all SAM-2 parameters remain frozen to ensure gradient updates exclusively modify the U-Net. Concurrently, the initial grayscale image $i$ is converted to RGB and fed to SAM-2's image encoder. Finally, the mask decoder integrates outputs from both the prompt encoder (processing the logits) and the image encoder to generate the segmentation output $\hat{m} = \mathcal{S}(z) \in \mathbb{R}^{240 \times 240}$. \\

\noindent \textbf{Learning Objectives}
Consider a batch of flattened image pixels indexed by $v = 1,\dots,N$, where $z_u^v$ denotes the model's raw logits, $m_v \in \{0,1\}$ the ground-truth binary labels, and $p_v = \sigma(z_u^v) = (1 + e^{-z_u^v})^{-1}$ the predicted probabilities via sigmoid activation. We optimize the following composite loss:

\begin{equation}
\mathcal{L} = w_{\text{F}} \mathcal{L}_{\text{F}} + w_{\text{D}} \mathcal{L}_{\text{D}} + w_{\text{LCD}} \mathcal{L}_{\text{LCD}}
\end{equation}

\noindent The \textbf{focal loss} $\mathcal{L}_{\text{F}}$ \cite{DBLP:journals/corr/abs-1708-02002} mitigates class imbalance by emphasizing hard examples (small liver area). For each pixel $v$, we define the class-adjusted probability $p_t^v$ and loss component $\ell_v^{\text{F}}$ as:
\begin{equation}
    p_t^v = \begin{cases} 
    p_v & \text{if } m_v = 1 \\
    1 - p_v & \text{if } m_v = 0 
\end{cases}, \quad
\ell_v^{\text{F}} = -\alpha_t (1 - p_t^v)^\gamma \log(p_t^v)
\end{equation}
where $\alpha_t$ is the class balancing factor and $\gamma \geq 0$ is the focusing parameter. The aggregate focal loss is given by:
\begin{equation}
   \mathcal{L}_{\text{F}} = \frac{1}{N} \sum_{v=1}^N \ell_v^{\text{F}} 
\end{equation}

\noindent The \textbf{dice loss} $\mathcal{L}_{\text{D}}$, which measures region overlap using the dice similarity coefficient (DS), is augmented with the \textbf{log-cosh dice loss} $\mathcal{L}_{\text{LCD}}$ \cite{Jadon_2020} to stabilize gradients during optimization:
\begin{equation}
    \mathcal{L}_{\text{LCD}} = \log \left( \cosh \left( \mathcal{L}_{\text{D}} \right) \right)
\end{equation}

\subsection{From 2D Segmentation Masks to 3D Volume}
Clinicians require a coherent 3D volume reconstructed from segmented 2D slices for preoperative planning. To generate volumetric representations from segmented slices, an affine transformation is applied to define spatial spacing and world coordinates. When a sufficient number of slices is available, this affine transform is applied directly to stacked segmentations. However, with sparse slices (common in MR), we could first enhance resolution via interpolation or super-resolution techniques \cite{greenspan2009super,isaac2015super} before applying the affine transformation.

\section{Experiments}
\subsection{Datasets and Implementation Details}
To evaluate the pipeline's effectiveness, we used the CHAOS (Combined (CT-MR) Healthy Abdominal Organ Segmentation) dataset \cite{CHAOSdata2019}. This dataset comprises 40 volumetric scans: 20 CT and 20 MR acquisitions, including both T2-SPIR (Spectral Presaturation with Inversion Recovery) and T1-DUAL (In-Phase/Out-Phase) sequences. This multi-modal composition introduces significant variations in noise patterns, contrast, intensity, and resolution across training, validation, and test sets, enabling robust testing under realistic conditions.

We used the ViT-H SAM2 model, since GPU limitations (NVIDIA RTX 2080 Ti w/ 12GB of RAM) allowed it. All edge detection methods were implemented using the Kornia \cite{eriba2019kornia} library, which is a differentiable computer vision library. The loss function weights were empirically chosen as follows: $w_F = 2$, $w_D = 2$ and $w_{LCD} = 3$. The model was trained for 250 epochs with a patience counter of 50, using a constant learning rate of $10^{-3}$, and a batch size of 8.

\subsection{Evaluation Protocol}
We partitioned the data into three subsets: a training set ($D_T^{TRAIN}$) containing 20 volumes (10 CT + 10 MR), a validation set of 10 volumes (5 CT + 5 MR), and a test set ($D_T^{TEST}$) of 10 volumes (5 CT + 5 MR). To rigorously assess model robustness, we implement a two-phase training strategy. First, all models were trained on the full $D_T^{TRAIN}$ and evaluated on $D_T^{TEST}$. Subsequently, models are retrained on a minimal subset $D_J^{TRAIN} \subseteq D_T^{TRAIN}$ (containing only 1 CT and 1 MR volume) and tested on an expanded set $D_J^{TEST}$ (14 CT + 14 MR volumes), simulating data-scarce scenarios. 

During preprocessing, non-liver labels in the original MR annotations were masked to focus segmentation exclusively on liver structures. Furthermore, we conducted cross-modality generalization tests: models trained solely on T1-DUAL OOP MR data were evaluated on CT images, and conversely, models trained exclusively on CT were evaluated on MR (T1-DUAL OOP) data. This protocol explicitly evaluates OOD performance.

\section{Results}
\subsection{Quantitative Validation}
We used both DICE and IoU scores as our evaluation metrics, and we call mDice the mean Dice score over both in-distribution and out-of-distribution tasks. We benchmark our method against three U-Net baselines: imU-Net (using raw images), emU-Net (using edge maps), and sU-Net (using raw images and SAM-2). We also compare our model against another self-prompting method for SAM-2 \cite{wu2023self}, which we refer to as spSAM. spSAM uses point- and bounding box prompts sampled within the geometry of a coarse mask generated by a logistic regression function. This design enables comparison with standard deep-learning-based segmentation approaches while isolating the contributions of individual pipeline components. The results are shown in \textbf{Tables \ref{tab1} and \ref{tab2}}. As shown in \textbf{Table \ref{tab1}}, our model maintains competitive performance on combined datasets (\textbf{95.1 \%} DICE) while benefiting from SAM's refinement capabilities in data-scarce scenarios. Crucially, \textbf{Table \ref{tab2}} demonstrates our approach's superior OOD generalization, with substantially improved cross-modality segmentation performance compared to single-modality baselines (\textbf{+27.4\%} mDICE). Additionally, we also demonstrated our pipeline's superior segmentation performance in comparison to another method involving the self-prompting of a SAM-2 module on both ID (\textbf{+14.4\%} mDICE) and OOD segmentation tasks (\textbf{+9.1\%} mDICE).

\begin{table}[h]
  \centering
  \setlength{\tabcolsep}{4pt} % Column padding
  \renewcommand{\arraystretch}{1.1} % Row height 
  \caption{Performance comparison using Dice \& IoU scores (mean $\pm$ std): Models trained on full mixed dataset $D_T^{TRAIN}$ and tested on $D_T^{TEST}$ versus models trained on minimal mixed subset $D_J^{TRAIN}$ and tested on $D_J^{TEST}$} \label{tab1}
  \begin{tabular}{|c|cc|cc|}
    \hline
    \multirow{2}{*}{\bfseries Models} & \multicolumn{2}{c|}{\bfseries {$D_T^{TEST}$}} & \multicolumn{2}{c|}{\bfseries $D_J^{TEST}$} \\ 
    \cline{2-5} % Span columns 2-5
    & Dice\%$\uparrow$ & IoU\%$\uparrow$ & Dice\%$\uparrow$ & IoU\%$\uparrow$ \\
    \hline  
    imU-Net & 96.3 ± 3.1 & 93.0 ± 5.4 & 80.7 ± 24.3 & 72.9 ± 26.1 \\
    emU-Net & 94.7 ± 4.5 & 90.3 ± 7.4 & 78.8 ± 19.8 & 68.5 ± 22.1 \\
    sU-Net & 96.0 ± 3.7 & 92.5 ± 6.1 & 84.7 ± 23.2 & 78.3 ± 24.4 \\
    spSAM & 78.8 ± 17.3 & 67.7 ± 17.2 & 76.7 ± 21.7 & 66.4 ± 21.4 \\
    \textbf{Ours} & 95.1 ± 4.3 & 90.8 ± 7.3 & 83.6 ± 16.1 & 74.5 ± 20.1 \\
    \hline
  \end{tabular}
\end{table}

\begin{table}[h]
  \centering
  \setlength{\tabcolsep}{4pt} % Column padding
  \renewcommand{\arraystretch}{1.1} % Row height
  \caption{Cross-modality generalization performance (Dice \& IoU, mean $\pm$ std): Models trained on CT or MR data and evaluated on ID and OOD test sets.}
  \label{tab2}
  \begin{tabular}{|cc|cc|cc|}
    \hline
    \multicolumn{2}{|c|}{\multirow{2}{*}{\bfseries Models}} & \multicolumn{2}{c|}{\bfseries In-Distribution}  & \multicolumn{2}{c|}{\bfseries Out-of-Distribution} \\
    \cline{3-6}
    \multicolumn{2}{|c|}{} & Dice\%$\uparrow$ & IoU\%$\uparrow$ & Dice\%$\uparrow$ & IoU\%$\uparrow$      \\
    \hline
    \multirow{2}{*}{imU-Net} 
      & CT & \textbf{95.9 ± 2.3} & \textbf{93.1 ± 4.1} & 62.2 ± 28.3 & 51.1 ± 29.2 \\
      & MR & 93.3 ± 2.6 & 88.4 ± 4.6 & 55.9 ± 30.1 & 44.5 ± 27.8 \\
    \hline
    \multirow{2}{*}{emU-Net} 
      & CT & 94.4 ± 6.0 & 90.6 ± 8.2 & 80.2 ± 13.3 & 68.3 ± 17.6 \\
      & MR & 91.9 ± 3.9 & 86.0 ± 6.4 & 75.2 ± 21.3 & 64.1 ± 22.2 \\
    \hline
    \multirow{2}{*}{sU-Net} 
      & CT & 95.2 ± 2.5 & 91.8 ± 4.4 & 62.3 ± 32.7 & 52.9 ± 32.4 \\
      & MR & \textbf{93.7 ± 2.7} & \textbf{89.0 ± 4.7} & 37.6 ± 33.1 & 28.7 ± 27.6  \\
    \hline
    \multirow{2}{*}{spSAM} 
      & CT & 77.8 ± 20.7 & 67.5 ± 20.7 & 81.6 ± 13.2 & 70.0 ± 13.2 \\
      & MR & 81.1 ± 9.3 & 72.1 ± 9.2 & 73.1 ± 24.9 & 62.6 ± 24.9 \\
    \hline
    \multirow{2}{*}{\textbf{Ours}} 
      & CT & 95.1 ± 2.7 & 91.7 ± 4.8 & \textbf{87.6 ± 8.7} & \textbf{79.3 ± 12.8} \\
      & MR & 92.8 ± 3.6 & 87.5 ± 6.1 & \textbf{85.3 ± 16.3} & \textbf{77.0 ± 18.2} \\
    \hline
  \end{tabular}
\end{table}

\subsection{Qualitative Validation}
\begin{figure}[h]
    \centering
    \begin{subfigure}{0.48\textwidth}
        \centering
        \includegraphics[width=\linewidth]{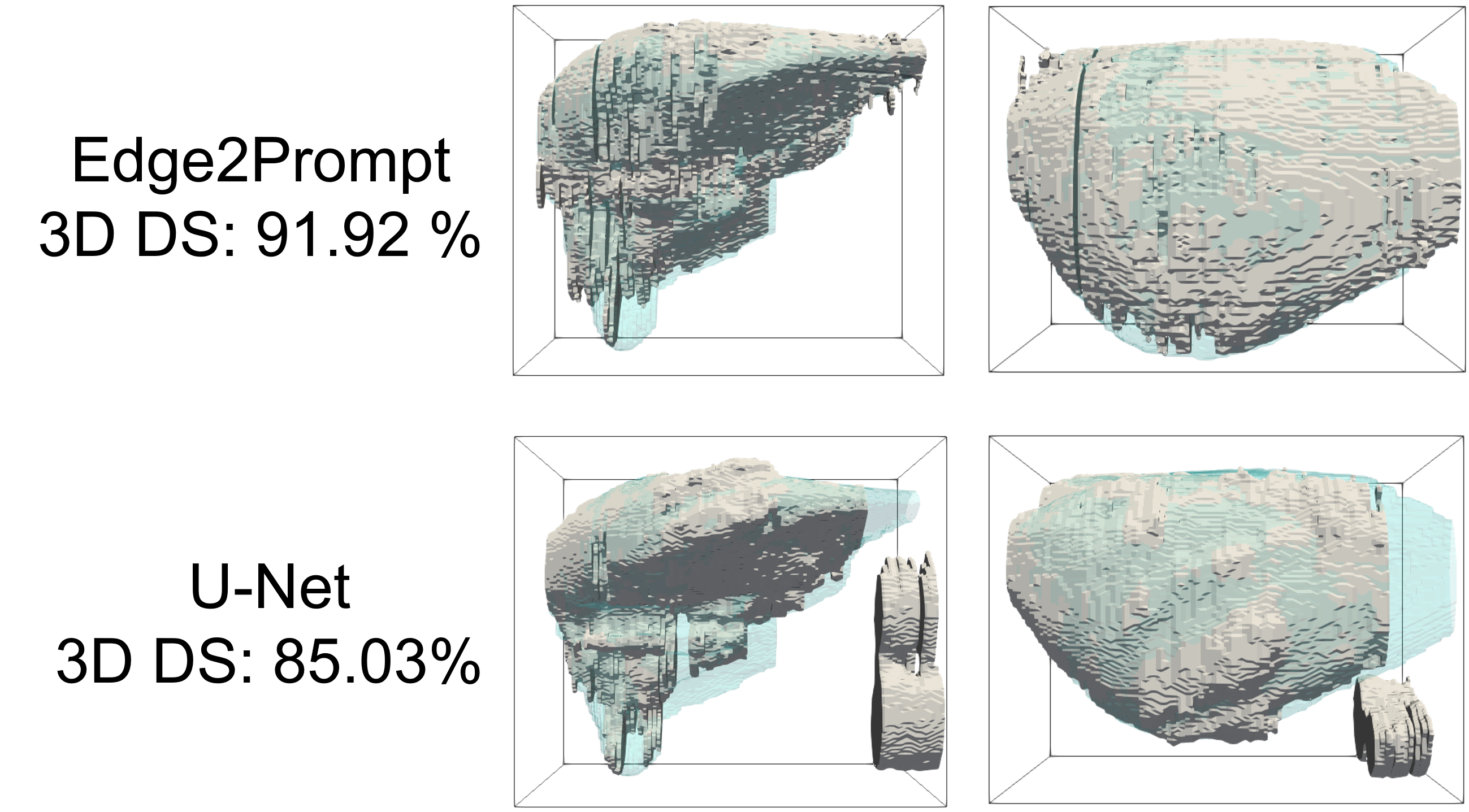}
        \caption{Best performing baseline volume}
        \label{fig:3dvisuA}
    \end{subfigure}
    \hfill
    \begin{subfigure}{0.48\textwidth}
        \centering
        \includegraphics[width=\linewidth]{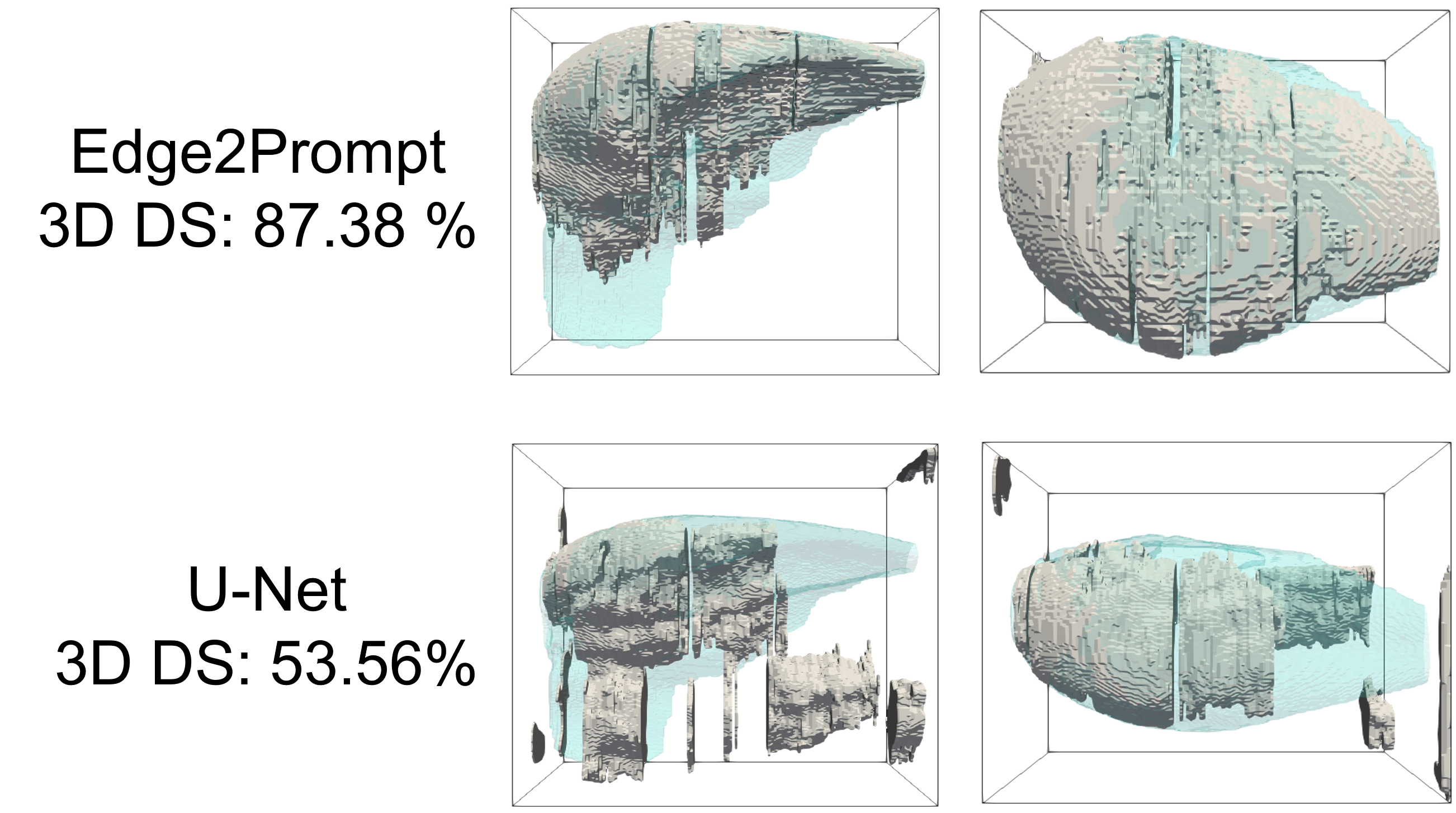}
        \caption{Poorest performing baseline volume}
        \label{fig:3dvisuB}
    \end{subfigure}
    \caption{Qualitative comparison of 3D CT volume reconstructions. Comparison of volume reconstructions after 2D segmentation using: Baseline U-Net, and \textbf{Edge2Prompt} (trained only on MR T1-DUAL OOP). Predicted volumes are overlaid on \textbf{the green ground truth} in frontal and axial views.}
    \label{fig:3dvisu}
\end{figure}

\textbf{Fig. \ref{fig:3dvisu}} qualitatively compares 3D CT volume reconstructions under OOD conditions, comparing segmentations from a baseline U-Net with our proposed method (both trained on MR data). Visual assessment and 3D Dice scores of two representative cases, one optimal and one poor for the baseline, demonstrate our model's superior ability to produce high-quality 3D reconstructions.

\subsection{Ablation study}
To demonstrate the necessity of each loss component, we conducted an ablation study on the mixed loss function using the cross-modality generalization task (training on MR T1-DUAL OOP images with \textbf{Edge2Prompt}). We trained and evaluated four distinct loss configurations. \textbf{Table \ref{tab:loss_ablation}} shows the optimal performance when combining all three components, yielding higher Dice scores and lower standard deviation. Furthermore, we evaluated model robustness by replacing the Canny edge detector with other edge detectors such as Sobel and Laplacian \cite{vincent2009descriptive} under identical task conditions. As shown in \textbf{Table \ref{tab:edge_ablation}}, while Canny achieves slightly superior performance, Sobel yields comparable results. 

% Combined table with common caption
\begin{table}
  \centering
  \caption{Results of ablation studies: (a) loss function components; (b) edge extraction techniques. Both show Dice \& IoU scores (mean $\pm$ std) from models trained on MR and tested on CT images (OOD).}
  \label{tab:combined}
  \begin{subtable}{0.45\textwidth}
    \centering
    \setlength{\tabcolsep}{1.5pt}
    \renewcommand{\arraystretch}{1.1}
    \caption{\textbf{Loss function ablation}}
    \label{tab:loss_ablation}
    \begin{tabular}{|c|cc|}
      \hline
      \multirow{2}{*}{\bfseries Loss} & \multicolumn{2}{c|}{\bfseries \textbf{OOD Results}} \\ 
      \cline{2-3}
      & Dice\%$\uparrow$ & IoU\%$\uparrow$ \\
      \hline  
      $\mathcal{L}_{F}$ & 70.2 ± 33.9 & 62.2 ± 32.5 \\
      $\mathcal{L}_{D}$ & 80.3 ± 23.3 & 73.3 ± 24.6  \\
      $\mathcal{L}_{LCD}$ & 73.6 ± 26.2 & 64.5 ± 26.5 \\
      $\mathcal{L}_{F}$ + $\mathcal{L}_{D}$& 82.7 ± 19.8 & 74.3 ± 22.5 \\
      $\mathcal{L}_{F}$ + $\mathcal{L}_{LCD}$& 83.1 ± 21.2 & 76.0 ± 22.2  \\
      $\mathcal{L}_{D}$ + $\mathcal{L}_{LCD}$ & 83.4 ± 16.5 & 75.1 ± 18.5 \\
      $\mathcal{L}$ & \textbf{85.3 ± 16.3} & \textbf{77.0 ± 18.3} \\
      \hline
    \end{tabular}
  \end{subtable}
  \hfill
  \begin{subtable}{0.45\textwidth}
    \centering
    \setlength{\tabcolsep}{1.5pt}
    \renewcommand{\arraystretch}{1.1}
    \caption{\textbf{Edge detectors ablation}}
    \label{tab:edge_ablation}
    \begin{tabular}{|c|cc|}
      \hline
      \multirow{2}{*}{\bfseries ED} & \multicolumn{2}{c|}{\bfseries \textbf{OOD Results}} \\ 
      \cline{2-3}
      & Dice\%$\uparrow$ & IoU\%$\uparrow$ \\
      \hline  
      Laplacian & 69.6 ± 37.2 & 55.5 ± 34.9 \\
      Sobel & 85.2 ± 17.2 & 76.6 ± 18.6  \\
      Canny & \textbf{85.3 ± 16.3} & \textbf{77.0 ± 18.3} \\
      \hline
    \end{tabular}
  \end{subtable}
\end{table}

\section{Discussion and Conclusion}
We present a simple yet effective solution to core challenges in medical image segmentation: domain dependence and data scarcity. Leveraging conventional image segmentation techniques (edge-prompted U-Net) with large-scale vision foundation models enables high-quality, self-prompted, modality-agnostic segmentation. Our pipeline matches the performance of classical segmentation models on conventional in-distribution tasks while outperforming them in fully out-of-distribution tasks and data-scarce settings. Complementing these results, we also compared our method to an alternative self-prompting approach using point/box prompts to condition SAM-2. Our pipeline again demonstrated superior performance on both in-distribution and out-of-distribution tasks. Furthermore, qualitative analysis based on reconstructed volumes underscores the method's clinical utility. Future research should optimize edge detection parameters and the loss function. Exploring alternative modality-agnostic feature extraction methods to improve U-Net input quality, alongside potential architectural improvements, would enhance segmentation. Validating the method on additional modalities like ultrasound and fluoroscopy would further verify robustness. Finally, developing methods to fine-tune SAM-2 without inducing domain dependence would significantly boost segmentation quality by mitigating limitations like object size sensitivity (causing high metric variance) and lack of medical image training.

\begin{comment}  %% removed for anonymized MICCAI 2025 submission.
    
    % The following acknowledgement and disclaimer sections should be removed for the double-blind review process.  
    % If and when your paper is accepted, reinsert the acknowledgement and the disclaimer clause in your final camera-ready version.

\begin{credits}
\subsubsection{\ackname} A bold run-in heading in small font size at the end of the paper is
used for general acknowledgments, for example: This study was funded
by X (grant number Y).

\subsubsection{\discintname}
It is now necessary to declare any competing interests or to specifically
state that the authors have no competing interests. Please place the
statement with a bold run-in heading in small font size beneath the
(optional) acknowledgments\footnote{If EquinOCS, our proceedings submission
system, is used, then the disclaimer can be provided directly in the system.},
for example: The authors have no competing interests to declare that are
relevant to the content of this article. Or: Author A has received research
grants from Company W. Author B has received a speaker honorarium from
Company X and owns stock in Company Y. Author C is a member of committee Z.
\end{credits}

\end{comment}
\bibliographystyle{ieeetr}
\bibliography{mybibliography}


%\end{thebibliography}
\end{document}